\documentclass[preprint,showpacs,preprintnumbers,amsmath,amssymb,nofootinbib]{revtex4-2}

\usepackage{graphicx}
\usepackage{dcolumn}
\usepackage{bm}
\usepackage{color}

\begin{document}

\newif\ifplot
\plottrue
\newcommand{\RR}[1]{[#1]}
\newcommand{\intsum}{\sum \kern -15pt \int}
\newfont{\Yfont}{cmti10 scaled 2074}
\newcommand{\Y}{\hbox{{\Yfont y}\phantom.}}
\def\O{{\cal O}}
\newcommand{\bra}[1]{\left< #1 \right| }
\newcommand{\braa}[1]{\left. \left< #1 \right| \right| }
\def\Bra#1#2{{\mbox{\vphantom{$\left< #2 \right|$}}}_{#1}
\kern -2.5pt \left< #2 \right| }
\def\Braa#1#2{{\mbox{\vphantom{$\left< #2 \right|$}}}_{#1}
\kern -2.5pt \left. \left< #2 \right| \right| }
\newcommand{\ket}[1]{\left| #1 \right> }
\newcommand{\kett}[1]{\left| \left| #1 \right> \right.}
\newcommand{\scal}[2]{\left< #1 \left| \mbox{\vphantom{$\left< #1 #2 \right|$}}
\right. #2 \right> }
\def\Scal#1#2#3{{\mbox{\vphantom{$\left<#2#3\right|$}}}_{#1}
{\left< #2 \left| \mbox{\vphantom{$\left<#2#3\right|$}}
\right. #3 \right> }}

\title{
  Investigation of interaction of circularly and linearly
  polarized photon beams with polarized $^3$He target}

\author{H.\ Wita{\l}a$^{1}$}
\email{henryk.witala@uj.edu.pl}
\author{J.\ Golak$^{1}$}
\author{R.\ Skibi\'nski$^{1}$}
\author{K.\ Topolnicki$^{1}$}
\author{V.\ Urbanevych$^{1}$}
\affiliation{
$^{1}$M. Smoluchowski Institute of Physics, Jagiellonian University,
PL-30348 Krak\'ow, Poland}

\date{\today}

\begin{abstract}
  Photodisintegration of polarized $^3$He by linearly or circularily
  polarized photons offers a rich choice of observables
  which can be calculated with high precision using a rigorous scheme  of
  three-nucleon Faddeev equations.
  Using the (semi)phenomenological  AV18 nucleon-nucleon potential 
  combined with the Urbana IX three-nucleon force we investigate sensitivity
  of $^3$He photodisintegration observables to underlying currents
  taken in the form of a single-nucleon current
  supplemented by two-body contributions for $\pi$- and $\rho$-meson
  exchanges  or incorporated  by the Siegert theorem. 
  Promising observables to be measured for two- and three-body
  fragmentation of $^3$He are identified. These observables form a challenging 
test ground  for consistent  forces and currents being under derivation 
  within the framework of chiral perturbation theory. 
For thre-body $^3$He photodisintegration several kinematicaly complete
configurations, including SST and FSI, are also discussed.
 
\end{abstract}


\maketitle

\section{Introduction}

Application of effective field theoretical methods in the form of
chiral perturbation theory (ChPT) puts investigation of 
properties of nuclear systems and their reactions on a new qualitative and
quantitative level. Derivation of consistent two- and many-body internucleonic
forces, and particularly the construction of nucleon-nucleon
(NN) \cite{epel_nn_n3lo,epel2006,machpr,new1,epel2}
and three-nucleon (3N) \cite{epel2002,3nf_n3lo_long,3nf_n3lo_short}
 interactions up to a fourth  (N$^4$LO) order of the chiral
expansion, provided a well grounded
nuclear Hamiltonian,  allowing for investigations of the importance
of a three-nucleon force on a much
 more solid theoretical basis \cite{binder2016,binder2018,eepel2019,eepel2019new}.  

Progress in the development of numerical methods for nuclear structure calculations
together with the availability of rigorous treatments of few-nucleon reactions 
within the Faddeev scheme made it possible to apply chiral two- and three-nucleon
forces in nuclear structure and 3N reaction calculations, and, by comparison
 of theoretical predictions to data, to test the underlying dynamics.
In this respect, the 3N system for which a large
data base of high precision cross sections and spin observables is
 available (\cite{glo96,Ndreview,Stephan1,Sekiguchi1,Stephan2,Stephan3,Ciepal,Sekiguchi2,Sekiguchi3,Tavakoli,Dadkan} 
and references therein)
for elastic nucleon-deuteron (Nd) scattering and the deuteron breakup reaction is
especially important.

For the electromagnetic sector an additional ingredient, the electromagnetic current has to be included.
Investigations of its structure are less advanced due to experimental difficulties as well as 
due to the lack of a consistent treatment of forces and currents~\cite{Carlson}. 
 This situation has changed with the recent 
derivations of electro-weak currents within 
ChPT~\cite{skoll2009,skoll2011,krebs2017,krebs2019,pastore2008,pastore2009},
which are, at least partly, consistent up to a given order of
the chiral expansion to
 the corresponding chiral
nuclear interaction {\footnote{Note that 
despite the fact that the operator form of chiral currents is known
at the lowest orders,  
applications of chiral many-nucleon forces and currents
in practical 3N calculations
is presently questionable due to the lack of their consistent regularization.}}.
With the advent of consistent forces and currents the number 
of reactions in which details of chiral dynamics can be studied will increase significantly.
One of such reactions is photodisintegration of
$^3$He which can lead to two- and three-body fragmentation in the final state.
For energies of the incoming photons below the $\pi$-production threshold the
Faddeev scheme allows one to get numerically exact predictions. This,  
 together with the fact, that 
photodisintegration of $^3$He offers  numerous observables, seems to
predispose  these reactions to become a valuable tool to test the dynamics with
 its electromagnetic current operator ingredient.   

With the availability of low energy  high intensity
 polarized photon beams \cite{Kii2005,weller2008} 
 and polarized $^3$He targets \cite{kramer2007,seki2019} a
 set of observables offered by 
two- and three-body photodisintegration of $^3$He, which could be
measured with sufficient precision, is substantially
extended and includes, in addition to the unpolarized cross section, also 
analyzing powers and spin correlation coefficients.

It is the aim of the present study to investigate the sensitivity of observables 
in low energy $\vec {\gamma} + \vec {{^3}He}$
 photodisintegration to underlying currents.
 In anticipation of application of consistent chiral interactions and currents we  
 use the (semi)phenomenological AV18 \cite{AV18} nucleon-nucleon potential
  combined with the Urbana IX \cite{uIX}  three-nucleon force and
  the current operator 
 taken in the form of a single nucleon current
 supplemented by two-body  contributions either in the form
 of explicit $\pi$- and $\rho$-meson exchanges  
 or incorporated  by the Siegert theorem \cite{physrep_el}. We solve
  the 3N Faddeev equations for
  $^3$He photodisintegration \cite{physrep_el} with such dynamics 
and calculate all possible
  observables for two- and three-body fragmentation
  of polarized $^3$He induced with linearly or circularily polarized photons,
    investigating their sensitivity to the underlying current.

The paper is organized as follows: in Sec. \ref{form} we briefly  
describe the theoretical formalism  and derive expressions for
polarization observables in $^3$He photodisintegration  induced by
 circularily or linearly polarized photons interacting with a
polarized $^3$He.
We present and discuss  results for two- and three-body  fragmentation
of $^3$He in  Sec. \ref{results_2b} and summarize and
conclude in Sec. \ref{summary}.

\section{Theoretical formalism}
\label{form}

In the following we present briefly our treatment of $^3$He photodisintegration
within the Faddeev scheme. For details of the theoretical formalism and numerical
performance we refer the reader to \cite{physrep_el, Skibinski_3metods}.

The basic quantities from which all observables for photodisintegration of
$^3$He can be calculated are the nuclear matrix elements  
$N^{\lambda m}_{\{\mu_i\}}$ \cite{physrep_el}: 
\begin{eqnarray}
  N^{\lambda m}_{\{\mu_i\}} \equiv < \Psi^{(-)}_{f,\{\mu_i\}} \vert \vec j_{\lambda}
  \vert \Psi^{^3He}_m >
\label{eqa1.1}
\end{eqnarray}
of the $^3$He current operator $\vec j_{\lambda}$($\lambda=\pm 1$) between
the initial $^3$He state $\vert \Psi^{^3He}_m >$ with the spin projection $m$ onto the 
z-axis defined by the momentum of the incoming photon, and the final 3N scattering
state  $ \vert \Psi^{(-)}_{f,\{\mu_i\}} > $, which is either two-body
proton-deuteron (pd) or
three-body proton-proton-neutron (ppn) fragmentation, with
corresponding spin projections of the 
outgoing particles $\{\mu_i\}$.

Assuming that three nucleons interact with two- and three-nucleon forces, and 
writing down 3N Faddeev equations for the Faddeev components of
3N scattering states, allows us to 
express the nuclear matrix elements for two- and three-body fragmentation of
$^3$He in terms of an auxiliary
 state $\vert \tilde U >$, which fulfills the following 
Faddeev-like equation \cite{physrep_el}: 
\begin{eqnarray}
  \vert \tilde U > &=& [tG_0 + \frac {1} {2} (1+P)V^{(1)}G_0(1+tG_0)] (1+P)
  \vec j_{\lambda}   \vert  \Psi^{^3He}_m > \cr
  &&+  [tG_0P + \frac {1} {2} (1+P)V^{(1)}G_0(1+tG_0)P]
   \vert \tilde U > ~.
\label{eqa2.1}
\end{eqnarray}
The permutation operator $P \equiv P_{12}P_{23} + P_{13}P_{23}$ is given by
interchanges $P_{ij}$ of nucleons $i$ and $j$ and the NN t-operator $t$ results from the employed NN
potential  through the two-body Lippmann-Schwinger equation. $V^{(1)}$ is a part
of a 3N force operator $V_{123}=V^{(1)}+V^{(2)}+V^{(3)}$, which is
 symmetrical under  exchange of nucleons 2 and 3.

Using the state   $  \vert \tilde U > $
the nuclear matrix element for pd fragmentation of $^3$He is given
 by \cite{physrep_el}:
\begin{eqnarray}
  N^{\lambda m}_{\mu_p \mu_d} &=& <\phi_q \vert (1+P)\vec j_{\lambda}
\vert  \Psi^{^3He}_m > +  <\phi_q \vert P \vert \tilde U > ~,
\label{eqa3.1}
\end{eqnarray}  
with $\vert \phi_q > \equiv \vert \varphi_d \mu_d > \vert \vec q_p \mu_p >$, where
$\vert \varphi_d >$ is the internal deuteron state, $\mu_d$ and $\mu_p$ are the deuteron and
proton spin projections, respectively, and $\vec q_p$ is a relative proton-deuteron
momentum.

For three-body breakup the nuclear matrix element is given by \cite{physrep_el}:
\begin{eqnarray}
  N^{\lambda m}_{\mu_1 \mu_2 \mu_3} &=& <\varphi_0 \vert (1+tG_0)(1+P)\vec j_{\lambda}
\vert  \Psi^{^3He}_m > +  <\varphi_0 \vert (1+tG_0)P \vert  \tilde U > ~,
\label{eqa4.1}
\end{eqnarray}
with $\vert \varphi_0 > \equiv (1-P_{23}) \vert \vec p \mu_2 \mu_3 >
\vert \vec q \mu_1 >$, where $\mu_i$ ($i=1$ ,$2$, $3$) are nucleons' spin
projections, and $\vec p$, $\vec q$ are standard Jacobi momenta.

The density matrix of polarized photons written in the spherical basis
(rows labelled with $\lambda=+1$ and $-1$) is given by \cite{arenh2008}:
\begin{eqnarray}
  \rho^{\gamma} = \frac {1} {2}
  \left( \begin{array} {cc}
    1+P^{\gamma}_c       & P^{\gamma}_x + iP^{\gamma}_y \\
    P^{\gamma}_x - iP^{\gamma}_y    & 1-P^{\gamma}_c
\end{array} \right)
\label{eq1.1}
\end{eqnarray}
where  $P^{\gamma}_{x}$ ($P^{\gamma}_{y}$) is a linear and $P^{\gamma}_c$ a circular
 polarization of photon.

The density matrix of polarized $^3$He
(rows labelled with the $^3$He spin projection
 $m=+\frac {1} {2}$ and $m=-\frac {1} {2}$),   
with a polarization vector
$\vec P=(P_x,P_y,P_z)$ is given by:
\begin{eqnarray}
  \rho^{^3He} = \frac {1} {2}
  \left( \begin{array} {cc}
    1+P_z       &P_x + iP_y \\
    P_x - iP_y    & 1-P_z
\end{array} \right)
\label{eq1.2}
\end{eqnarray}

For the initial channel $\vec {\gamma} + \vec {^3He}$ with the z-axis taken
 in the direction of the 
incoming photon beam, the incoming state density matrix is a direct product
of the photon and $^3$He density matrices:
\begin{eqnarray}
  \rho^{in}=\rho^{\gamma} \otimes \rho^{^3He} = \frac {1} {4} 
  \left( \begin{array} {cc}
    1+P^{\gamma}_c       & P^{\gamma}_x + iP^{\gamma}_y \\
    P^{\gamma}_x - iP^{\gamma}_y    & 1-P^{\gamma}_c
  \end{array} \right)
   \otimes
  \left( \begin{array} {cc}
    1+P_z       &P_x + iP_y \\
    P_x - iP_y    & 1-P_z
\end{array} \right)
\label{eq1.3}
\end{eqnarray}

The cross section with polarized photons and polarized $^3$He: 
$\sigma^{pol}= Tr(N\rho^{in}N^+)$, 
 with the full transition amplitude $N$ containing the nuclear 
matrix elements
$N^{\lambda m}_{\{\mu_i\}}$,  is given by:
\begin{eqnarray}
  \sigma^{pol} &=& \sigma^o( 1 + P^{\gamma}_c A^{\gamma}_c + P^{\gamma}_x A^{\gamma}_x
  + P^{\gamma}_y A^{\gamma}_y + P_x A^{^3He}_x + P_y A^{^3He}_y + P_z A^{^3He}_z \cr
  &+& P^{\gamma}_c P_z C^{\gamma,{^3}He}_{c,z} +  P^{\gamma}_c P_x C^{\gamma,{^3}He}_{c,x} +
  P^{\gamma}_c P_y C^{\gamma,{^3}He}_{c,y} +  P^{\gamma}_x P_z C^{\gamma,{^3}He}_{x,z} +
  P^{\gamma}_x P_x C^{\gamma,{^3}He}_{x,x} \cr
  &+& P^{\gamma}_x P_y C^{\gamma,{^3}He}_{x,y} 
   + P^{\gamma}_y P _z C^{\gamma,{^3}He}_{y,z} + P^{\gamma}_y P_x C^{\gamma,{^3}He}_{y,x} +
  P^{\gamma}_y P_y C^{\gamma,{^3}He}_{y,y} )
\label{eq1.5}
\end{eqnarray}
with photon analyzing powers: $A^{\gamma}_c$, $A^{\gamma}_x$, and $A^{\gamma}_y$,
$^3$He-analyzing powers: $A^{^3He}_x$, $A^{^3He}_y$, and $A^{^3He}_z$, and
 spin correlation
 coefficients: $C^{\gamma,{^3}He}_{c,x}$, $C^{\gamma,{^3}He}_{c,y}$,
 $C^{\gamma,{^3}He}_{c,z}$,
 $C^{\gamma,{^3}He}_{x,x}$, $C^{\gamma,{^3}He}_{x,y}$, $C^{\gamma,{^3}He}_{x,z}$,
$C^{\gamma,{^3}He}_{y,x}$, $C^{\gamma,{^3}He}_{y,y}$, and $C^{\gamma,{^3}He}_{y,z}$ given by:
\begin{eqnarray}
  A^{\gamma}_c &=& 
   \frac { \sum\limits_{ \{\mu_i\} m}   N^{+1 m}_{\{\mu_i\}}  N^{* +1 m}_{\{\mu_i\}}
            -  N^{-1 m}_{\{\mu_i\}}  N^{* -1 m}_{\{\mu_i\}} }
  {\sum\limits_{ \{\mu_i\} m \lambda} \vert N^{\lambda m}_{\{\mu_i\}} \vert^2  } \cr
  A^{\gamma}_x &=&
  \frac { \sum\limits_{ \{\mu_i\} m}   N^{+1 m}_{\{\mu_i\}}  N^{* -1 m}_{\{\mu_i\}}
            +  N^{-1 m}_{\{\mu_i\}}  N^{* +1 m}_{\{\mu_i\}} }
  {\sum\limits_{ \{\mu_i\} m \lambda} \vert N^{\lambda m}_{\{\mu_i\}} \vert^2  } 
  = \frac { \sum\limits_{ \{\mu_i\} m}
    2\Re \left[ N^{+1 m}_{\{\mu_i\}}  N^{* -1 m}_{\{\mu_i\}} \right] }
  {\sum\limits_{ \{\mu_i\} m \lambda} \vert N^{\lambda m}_{\{\mu_i\}} \vert^2  } \cr
  A^{\gamma}_y &=&
  \frac { \sum\limits_{ \{\mu_i\} m}  iN^{+1 m}_{\{\mu_i\}}  N^{* -1 m}_{\{\mu_i\}}
            - iN^{-1 m}_{\{\mu_i\}}  N^{* +1 m}_{\{\mu_i\}} }
  {\sum\limits_{ \{\mu_i\} m \lambda} \vert N^{\lambda m}_{\{\mu_i\}} \vert^2  } 
  = \frac { \sum\limits_{ \{\mu_i\} m}
    -2\Im \left[ N^{+1 m}_{\{\mu_i\}}  N^{* -1 m}_{\{\mu_i\}} \right] }
  {\sum\limits_{ \{\mu_i\} m \lambda} \vert N^{\lambda m}_{\{\mu_i\}} \vert^2  } \cr
  A^{{^3}He}_z &=&
  \frac { \sum\limits_{ \{\mu_i\} \lambda}   N^{\lambda \frac {1} {2}}_{\{\mu_i\}}
    N^{* \lambda \frac {1} {2}}_{\{\mu_i\}}
 -  N^{\lambda -\frac {1} {2}}_{\{\mu_i\}}  N^{* \lambda -\frac {1} {2}}_{\{\mu_i\}} }
  {\sum\limits_{ \{\mu_i\} m \lambda} \vert N^{\lambda m}_{\{\mu_i\}} \vert^2  } \cr
  A^{{^3}He}_x &=&
  \frac { \sum\limits_{ \{\mu_i\} \lambda}   N^{\lambda \frac {1} {2}}_{\{\mu_i\}}
    N^{* \lambda -\frac {1} {2}}_{\{\mu_i\}}
 +  N^{\lambda -\frac {1} {2}}_{\{\mu_i\}}  N^{* \lambda \frac {1} {2}}_{\{\mu_i\}} }
  {\sum\limits_{ \{\mu_i\} m \lambda} \vert N^{\lambda m}_{\{\mu_i\}} \vert^2  } 
  = \frac { \sum\limits_{ \{\mu_i\} \lambda}  2\Re \left[
      N^{\lambda \frac {1} {2}}_{\{\mu_i\}}  N^{* \lambda -\frac {1} {2}}_{\{\mu_i\}} \right] }
  {\sum\limits_{ \{\mu_i\} m \lambda} \vert N^{\lambda m}_{\{\mu_i\}} \vert^2  } \cr
  A^{{^3}He}_y &=&
  \frac { \sum\limits_{ \{\mu_i\} \lambda}   iN^{\lambda \frac {1} {2}}_{\{\mu_i\}}
    N^{* \lambda -\frac {1} {2}}_{\{\mu_i\}}
 -  iN^{\lambda -\frac {1} {2}}_{\{\mu_i\}}  N^{* \lambda \frac {1} {2}}_{\{\mu_i\}} }
  {\sum\limits_{ \{\mu_i\} m \lambda} \vert N^{\lambda m}_{\{\mu_i\}} \vert^2  }
  = \frac { \sum\limits_{ \{\mu_i\} \lambda}  -2\Im \left[
      N^{\lambda \frac {1} {2}}_{\{\mu_i\}}  N^{* \lambda -\frac {1} {2}}_{\{\mu_i\}} \right] }
  {\sum\limits_{ \{\mu_i\} m \lambda} \vert N^{\lambda m}_{\{\mu_i\}} \vert^2  } \cr
  C^{\gamma,{^3}He}_{c,z} &=&
  \frac { \sum\limits_{ \{\mu_i\}}   N^{+1 \frac {1} {2}}_{\{\mu_i\}}
    N^{* +1 \frac {1} {2}}_{\{\mu_i\}}
 +  N^{-1 -\frac {1} {2}}_{\{\mu_i\}} N^{* -1 -\frac {1} {2}}_{\{\mu_i\}}
 -  N^{+1 -\frac {1} {2}}_{\{\mu_i\}}  N^{* +1 -\frac {1} {2}}_{\{\mu_i\}} 
 -  N^{-1 \frac {1} {2}}_{\{\mu_i\}}  N^{* -1 \frac {1} {2}}_{\{\mu_i\}} } 
    {\sum\limits_{ \{\mu_i\} m \lambda} \vert N^{\lambda m}_{\{\mu_i\}} \vert^2  } \cr
    C^{\gamma,{^3}He}_{c,x} &=&
    \frac { \sum\limits_{ \{\mu_i\}}   N^{+1 \frac {1} {2}}_{\{\mu_i\}}
    N^{* +1 -\frac {1} {2}}_{\{\mu_i\}}
 +  N^{+1 -\frac {1} {2}}_{\{\mu_i\}} N^{* +1 \frac {1} {2}}_{\{\mu_i\}}
 -  N^{-1 \frac {1} {2}}_{\{\mu_i\}}  N^{* -1 -\frac {1} {2}}_{\{\mu_i\}} 
 -  N^{-1 -\frac {1} {2}}_{\{\mu_i\}}  N^{* -1 \frac {1} {2}}_{\{\mu_i\}} } 
    {\sum\limits_{ \{\mu_i\} m \lambda} \vert N^{\lambda m}_{\{\mu_i\}} \vert^2  } \cr
  &=& \frac { \sum\limits_{ \{\mu_i\}}  2\Re \left[ N^{+1 \frac {1} {2}}_{\{\mu_i\}}
    N^{* +1 -\frac {1} {2}}_{\{\mu_i\}}
 -  N^{-1 \frac {1} {2}}_{\{\mu_i\}}  N^{* -1 -\frac {1} {2}}_{\{\mu_i\}}\right] } 
    {\sum\limits_{ \{\mu_i\} m \lambda} \vert N^{\lambda m}_{\{\mu_i\}} \vert^2  } \cr
    C^{\gamma,{^3}He}_{c,y} &=&
    \frac { \sum\limits_{ \{\mu_i\}}   iN^{+1 \frac {1} {2}}_{\{\mu_i\}}
    N^{* +1 -\frac {1} {2}}_{\{\mu_i\}}
 +  iN^{-1 -\frac {1} {2}}_{\{\mu_i\}} N^{* -1 \frac {1} {2}}_{\{\mu_i\}}
 -  iN^{+1 -\frac {1} {2}}_{\{\mu_i\}}  N^{* +1 \frac {1} {2}}_{\{\mu_i\}} 
 -  iN^{-1 \frac {1} {2}}_{\{\mu_i\}}  N^{* -1 -\frac {1} {2}}_{\{\mu_i\}} } 
    {\sum\limits_{ \{\mu_i\} m \lambda} \vert N^{\lambda m}_{\{\mu_i\}} \vert^2  } \cr
  &=& \frac { \sum\limits_{ \{\mu_i\}}  -2\Im \left[ N^{+1 \frac {1} {2}}_{\{\mu_i\}}
    N^{* +1 -\frac {1} {2}}_{\{\mu_i\}}
 +  N^{-1 -\frac {1} {2}}_{\{\mu_i\}}  N^{* -1 \frac {1} {2}}_{\{\mu_i\}}\right] } 
    {\sum\limits_{ \{\mu_i\} m \lambda} \vert N^{\lambda m}_{\{\mu_i\}} \vert^2  } \cr
    C^{\gamma,{^3}He}_{x,z} &=&
    \frac { \sum\limits_{ \{\mu_i\}}   N^{+1 \frac {1} {2}}_{\{\mu_i\}}
    N^{* -1 \frac {1} {2}}_{\{\mu_i\}}
 +  N^{-1 \frac {1} {2}}_{\{\mu_i\}} N^{* +1 \frac {1} {2}}_{\{\mu_i\}}
 -  N^{+1 -\frac {1} {2}}_{\{\mu_i\}}  N^{* -1 -\frac {1} {2}}_{\{\mu_i\}} 
 -  N^{-1 -\frac {1} {2}}_{\{\mu_i\}}  N^{* +1 -\frac {1} {2}}_{\{\mu_i\}} } 
    {\sum\limits_{ \{\mu_i\} m \lambda} \vert N^{\lambda m}_{\{\mu_i\}} \vert^2  } \cr
  &=& \frac { \sum\limits_{ \{\mu_i\}}  2\Re \left[ N^{+1 \frac {1} {2}}_{\{\mu_i\}}
    N^{* -1 \frac {1} {2}}_{\{\mu_i\}}
 -  N^{+1 -\frac {1} {2}}_{\{\mu_i\}}  N^{* -1 -\frac {1} {2}}_{\{\mu_i\}}\right] } 
    {\sum\limits_{ \{\mu_i\} m \lambda} \vert N^{\lambda m}_{\{\mu_i\}} \vert^2  } \cr
    C^{\gamma,{^3}He}_{x,x} &=&
    \frac { \sum\limits_{ \{\mu_i\}}   N^{+1 \frac {1} {2}}_{\{\mu_i\}}
    N^{* -1 -\frac {1} {2}}_{\{\mu_i\}}
 +  N^{+1 -\frac {1} {2}}_{\{\mu_i\}} N^{* -1 \frac {1} {2}}_{\{\mu_i\}}
 +  N^{-1 \frac {1} {2}}_{\{\mu_i\}}  N^{* +1 -\frac {1} {2}}_{\{\mu_i\}} 
 +  N^{-1 -\frac {1} {2}}_{\{\mu_i\}}  N^{* +1 \frac {1} {2}}_{\{\mu_i\}} } 
    {\sum\limits_{ \{\mu_i\} m \lambda} \vert N^{\lambda m}_{\{\mu_i\}} \vert^2  } \cr
  &=& \frac { \sum\limits_{ \{\mu_i\}}  2\Re \left[ N^{+1 \frac {1} {2}}_{\{\mu_i\}}
    N^{* -1 -\frac {1} {2}}_{\{\mu_i\}}
 +  N^{+1 -\frac {1} {2}}_{\{\mu_i\}}  N^{* -1 \frac {1} {2}}_{\{\mu_i\}}\right] } 
    {\sum\limits_{ \{\mu_i\} m \lambda} \vert N^{\lambda m}_{\{\mu_i\}} \vert^2  } \cr
    C^{\gamma,{^3}He}_{x,y} &=&
    \frac { \sum\limits_{ \{\mu_i\}}   iN^{+1 \frac {1} {2}}_{\{\mu_i\}}
    N^{* -1 -\frac {1} {2}}_{\{\mu_i\}}
 -  iN^{+1 -\frac {1} {2}}_{\{\mu_i\}} N^{* -1 \frac {1} {2}}_{\{\mu_i\}}
 +  iN^{-1 \frac {1} {2}}_{\{\mu_i\}}  N^{* +1 -\frac {1} {2}}_{\{\mu_i\}} 
 -  iN^{-1 -\frac {1} {2}}_{\{\mu_i\}}  N^{* +1 \frac {1} {2}}_{\{\mu_i\}} } 
    {\sum\limits_{ \{\mu_i\} m \lambda} \vert N^{\lambda m}_{\{\mu_i\}} \vert^2  } \cr
  &=& \frac { \sum\limits_{ \{\mu_i\}}  -2\Im \left[ N^{+1 \frac {1} {2}}_{\{\mu_i\}}
    N^{* -1 -\frac {1} {2}}_{\{\mu_i\}}
 +  N^{-1 \frac {1} {2}}_{\{\mu_i\}}  N^{* +1 -\frac {1} {2}}_{\{\mu_i\}}\right] } 
    {\sum\limits_{ \{\mu_i\} m \lambda} \vert N^{\lambda m}_{\{\mu_i\}} \vert^2  } \cr
    C^{\gamma,{^3}He}_{y,z} &=&
    \frac { \sum\limits_{ \{\mu_i\}}   iN^{+1 \frac {1} {2}}_{\{\mu_i\}}
    N^{* -1 \frac {1} {2}}_{\{\mu_i\}}
 -  iN^{+1 -\frac {1} {2}}_{\{\mu_i\}} N^{* -1 -\frac {1} {2}}_{\{\mu_i\}}
 -  iN^{-1 \frac {1} {2}}_{\{\mu_i\}}  N^{* +1 \frac {1} {2}}_{\{\mu_i\}} 
 +  iN^{-1 -\frac {1} {2}}_{\{\mu_i\}}  N^{* +1 -\frac {1} {2}}_{\{\mu_i\}} } 
    {\sum\limits_{ \{\mu_i\} m \lambda} \vert N^{\lambda m}_{\{\mu_i\}} \vert^2  } \cr
  &=& \frac { \sum\limits_{ \{\mu_i\}}  -2\Im \left[ N^{+1 \frac {1} {2}}_{\{\mu_i\}}
    N^{* -1 \frac {1} {2}}_{\{\mu_i\}}
 +  N^{-1 -\frac {1} {2}}_{\{\mu_i\}}  N^{* +1 -\frac {1} {2}}_{\{\mu_i\}}\right] } 
    {\sum\limits_{ \{\mu_i\} m \lambda} \vert N^{\lambda m}_{\{\mu_i\}} \vert^2  }  \cr
    C^{\gamma,{^3}He}_{y,x} &=&
    \frac { \sum\limits_{ \{\mu_i\}}   iN^{+1 \frac {1} {2}}_{\{\mu_i\}}
    N^{* -1 -\frac {1} {2}}_{\{\mu_i\}}
 +  iN^{+1 -\frac {1} {2}}_{\{\mu_i\}} N^{* -1 \frac {1} {2}}_{\{\mu_i\}}
 -  iN^{-1 \frac {1} {2}}_{\{\mu_i\}}  N^{* +1 -\frac {1} {2}}_{\{\mu_i\}} 
 -  iN^{-1 -\frac {1} {2}}_{\{\mu_i\}}  N^{* +1 \frac {1} {2}}_{\{\mu_i\}} } 
    {\sum\limits_{ \{\mu_i\} m \lambda} \vert N^{\lambda m}_{\{\mu_i\}} \vert^2  } \cr
  &=& \frac { \sum\limits_{ \{\mu_i\}}  -2\Im \left[ N^{+1 \frac {1} {2}}_{\{\mu_i\}}
    N^{* -1 -\frac {1} {2}}_{\{\mu_i\}}
 +  N^{+1 -\frac {1} {2}}_{\{\mu_i\}}  N^{* -1 \frac {1} {2}}_{\{\mu_i\}}\right] } 
    {\sum\limits_{ \{\mu_i\} m \lambda} \vert N^{\lambda m}_{\{\mu_i\}} \vert^2  }  \cr
    C^{\gamma,{^3}He}_{y,y} &=&
    \frac { \sum\limits_{ \{\mu_i\}}   -N^{+1 \frac {1} {2}}_{\{\mu_i\}}
    N^{* -1 -\frac {1} {2}}_{\{\mu_i\}}
 +  N^{+1 -\frac {1} {2}}_{\{\mu_i\}} N^{* -1 \frac {1} {2}}_{\{\mu_i\}}
 +  N^{-1 \frac {1} {2}}_{\{\mu_i\}}  N^{* +1 -\frac {1} {2}}_{\{\mu_i\}} 
 -  N^{-1 -\frac {1} {2}}_{\{\mu_i\}}  N^{* +1 \frac {1} {2}}_{\{\mu_i\}} } 
    {\sum\limits_{ \{\mu_i\} m \lambda} \vert N^{\lambda m}_{\{\mu_i\}} \vert^2  } \cr
  &=& \frac { \sum\limits_{ \{\mu_i\}}  2\Re \left[ -N^{+1 \frac {1} {2}}_{\{\mu_i\}}
    N^{* -1 -\frac {1} {2}}_{\{\mu_i\}}
 +  N^{+1 -\frac {1} {2}}_{\{\mu_i\}}  N^{* -1 \frac {1} {2}}_{\{\mu_i\}}\right] } 
    {\sum\limits_{ \{\mu_i\} m \lambda} \vert N^{\lambda m}_{\{\mu_i\}} \vert^2  } 
\label{eq1.6}
\end{eqnarray}

Some of these observables vanish for two- as well as for three-body in plane
photodisintegration of $^3$He. Namely, assuming that outgoing particles move in
a plane, which  contains also  the photon beam incoming along z-axis, leads
to the following symmetry relation for the nuclear matrix elements
$N^{\lambda m}_{\{\mu_i\}}$:
\begin{eqnarray}
  N^{-\lambda -m}_{\{-\mu_i\}} = (-1)^{ (1 - m - \lambda + \sum \mu_i) }
  N^{\lambda m}_{\{\mu_i\}} ~.
\label{eq1.7}
\end{eqnarray}
Applying that relation to the polarization observables of Eq.~(\ref{eq1.6})
shows, that the only nonvanishing analyzing powers are $A^{\gamma}_x$ and
$A^{^3He}_y$, and the nonvanishing  spin correlation coefficients are:
$C^{\gamma,{^3}He}_{c,z}$,  $C^{\gamma,{^3}He}_{c,x}$, $C^{\gamma,{^3}He}_{x,y}$,
$C^{\gamma,{^3}He}_{y,z}$, and  $C^{\gamma,{^3}He}_{y,x}$.
The analyzing powers: $A^{\gamma}_c$,  $A^{\gamma}_y$, $A^{^3He}_x$, and $A^{^3He}_z$,
as well as spin correlation coefficients: $C^{\gamma,{^3}He}_{c,y}$,
$C^{\gamma,{^3}He}_{x,z}$, $C^{\gamma,{^3}He}_{x,x}$, and $C^{\gamma,{^3}He}_{y,y}$, vanish.

The same is true for kinematically incomplete three-body fragmentation with
 a reaction plane formed by the incoming photon beam and 
the momentum of the one nucleon detected in the final state.

\section{Results and discussion}
\label{results_2b}

To investigate the sensitivity of low energy $^3$He
 photodisintegration observables to
the underlying current operator we solved Faddeev-like Eq. (\ref{eqa2.1}) for
the state $\vert \tilde U >$ at three energies of the incoming photons: 
$E_{\gamma}=15$, $20$, and $30$~MeV, using our standard approach based on
momentum  space partial wave decomposition \cite{glo96,hub97,physrep_el,book}. 
The high precision (semi)phenomenological  nucleon-nucleon potential
AV18 \cite{AV18} together with the  Urbana IX \cite{uIX} three-nucleon force
 was used.  Including this 3N force reproduces the experimental $^3$He binding energy. 
The AV18 potential contains electromagnetic parts \cite{AV18}.
 They are all kept in our treatment of the $^3$He bound state \cite{nogga2003}
 but for the 3N continuum we keep only the strong forces.
  We solve the set of coupled integral equations in the two Jacobi
 variables by iteration, generating the
 multiple scattering series separately for each fixed total 3N system 
  angular momentum J and parity.
 We neglect the coupling of 
 states with total isospin $T = \frac {1} {2}$ and $T = \frac {3} {2}$,
 which is due to charge independence breaking for neutron-proton (np)  and
 proton-proton (pp) forces but keep both isospins T. The difference between
 pp and np forces is, however, taken into account by applying the
 ``$\frac {2} {3} - \frac {1} {3}$'' rule \cite{wit89,wit91}.

The single nucleon current was augmented by explicit $\pi$- and $\rho$-like
 two-body currents which fulfill the current continuity equation
 together with the corresponding parts of the AV18
  potential \cite{physrep_el}. As an
 alternative to explicit two-body contributions 
 we  employed  the Siegert theorem \cite{physrep_el}, which induces
 many-body contributions to the current operator.

\begin{figure}[htbp] 
\includegraphics[scale=0.7]{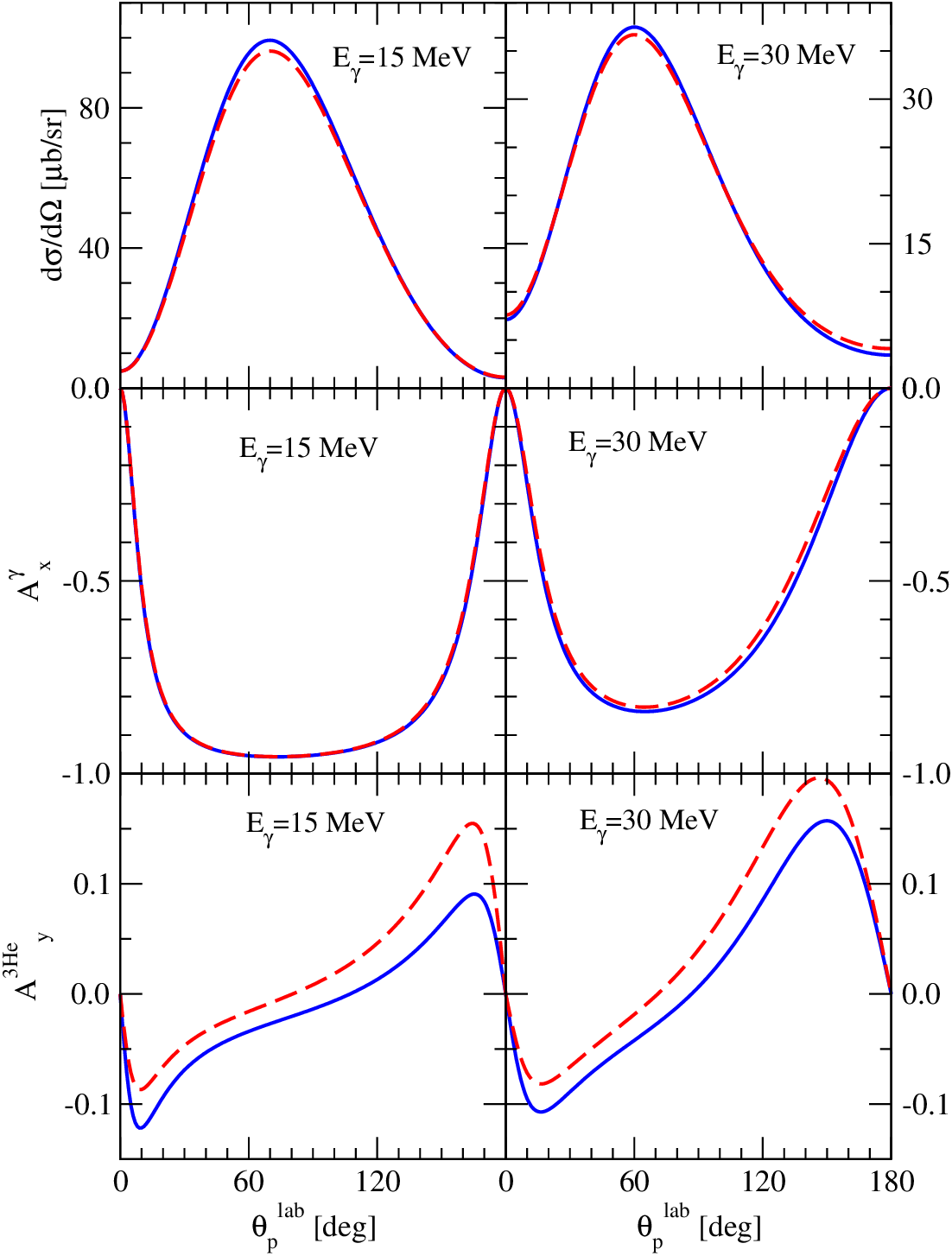}
\caption{(Color online) 
 The unpolarized cross section and the analyzing powers: $A^{\gamma}_x$ and 
  $A^{^3He}_y$, for two-body $^3$He
  photodisintegration $\vec {^3He}(\vec {\gamma},p)d$ at $E_{\gamma}=15$~MeV
(left column) and $E_{\gamma}=30$~MeV (right column).
Presented results are based on AV18 NN interaction combined with
Urbana IX 3NF, and $^3$He current which, in addition to single nucleon
current, contained two-body exchange contributions taken in the form of
meson-exchange currents ( (blue) solid line) or by Siegert theorem
( (red) dashed line).
\label{fig1}}
\end{figure}

\begin{figure}[htbp] 
\includegraphics[scale=0.64]{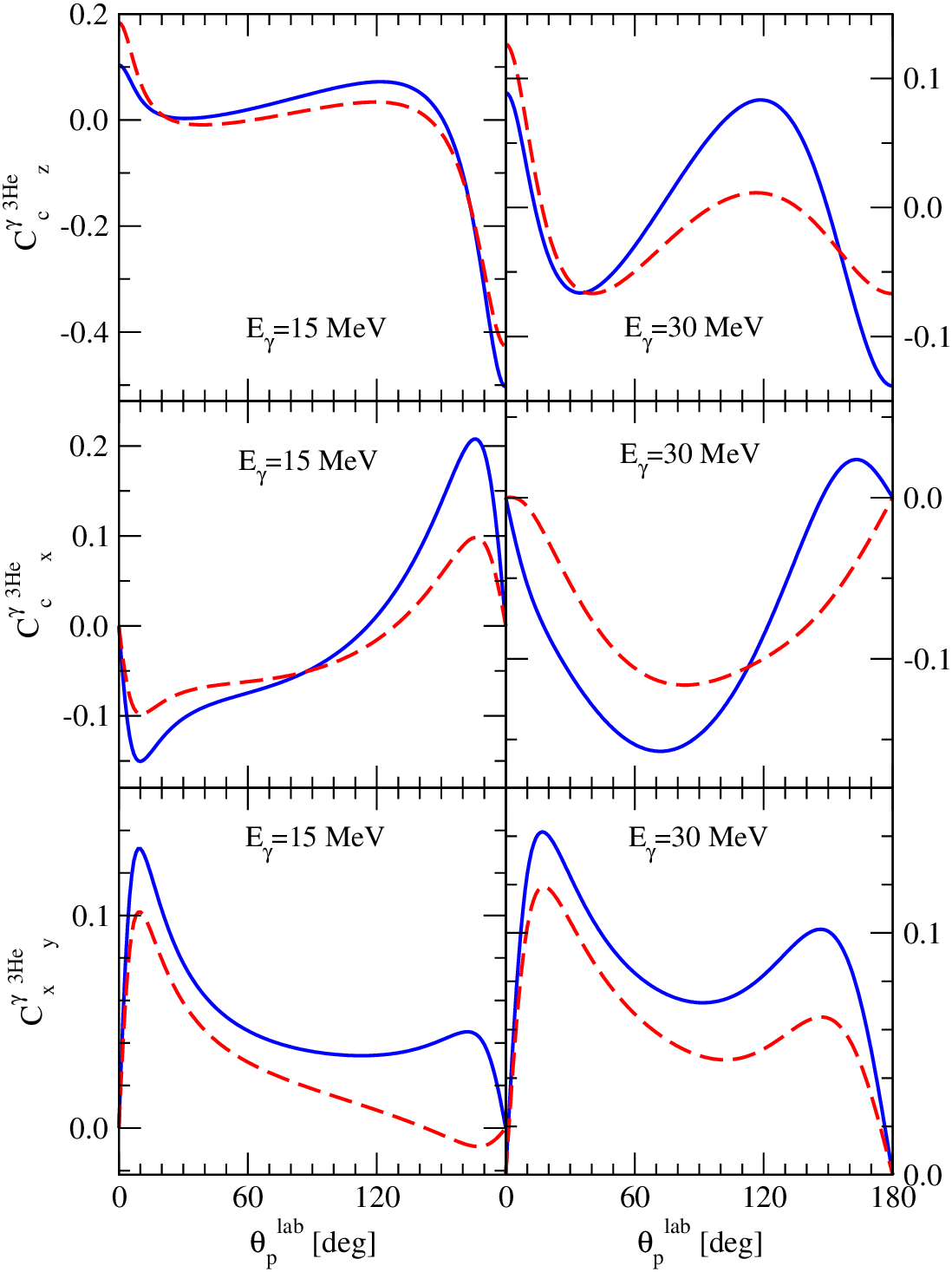}
\caption{(Color online) 
 The spin correlation coefficients $C^{\gamma-^3He}_{c-z}$, $C^{\gamma-^3He}_{c-x}$, and 
 $C^{\gamma-^3He}_{x-y}$  
  for two-body $^3$He
  photodisintegration $\vec {^3He}(\vec {\gamma},p)d$ at $E_{\gamma}=15$~MeV
(left column) and $E_{\gamma}=30$~MeV (right column).
Lines are the same as in Fig.\ref{fig1}.
\label{fig2}}
\end{figure}

\begin{figure}[htbp] 
\includegraphics[scale=0.7]{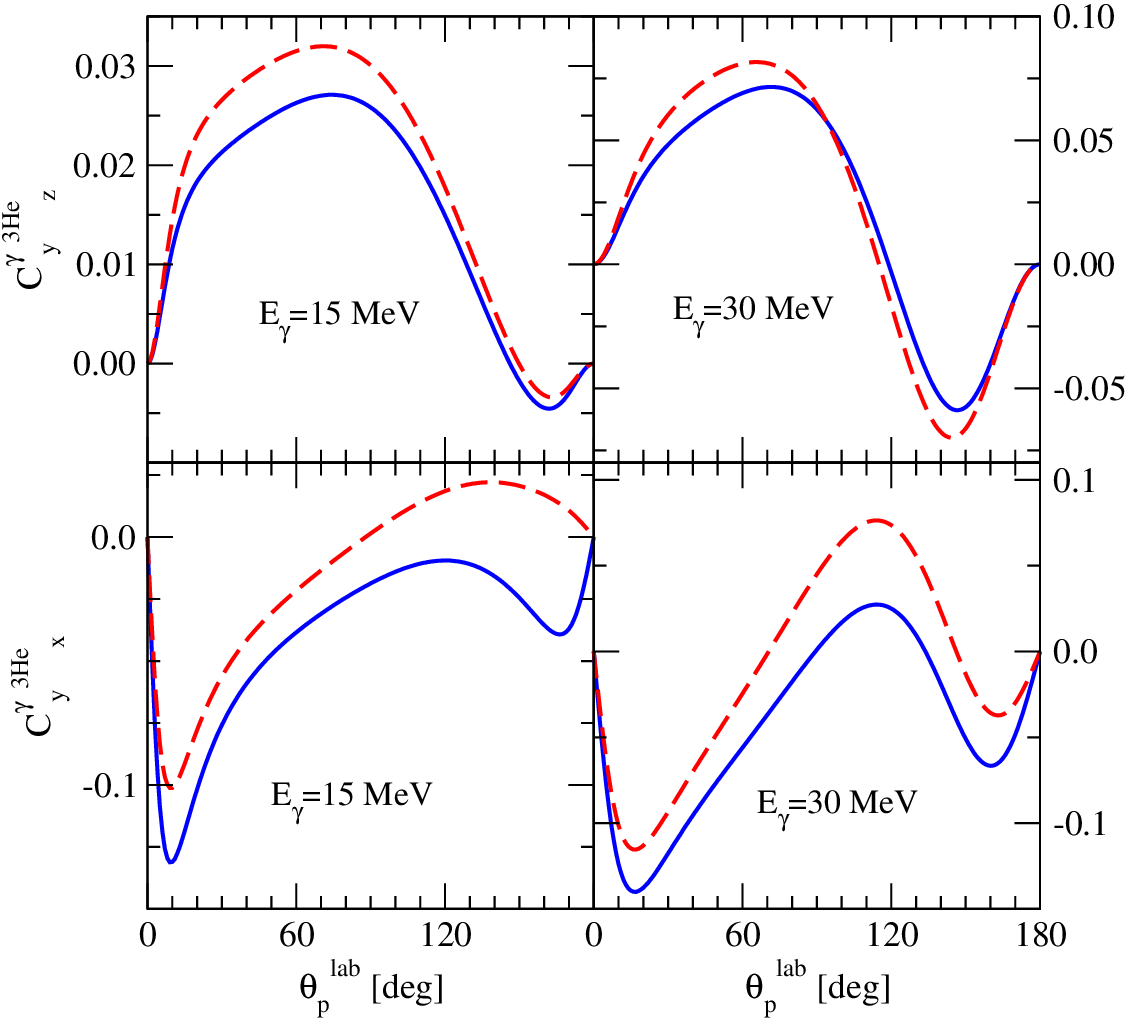}
\caption{(Color online) 
  The spin correlation coefficients $C^{\gamma~^3He}_{y~z}$ and  $C^{\gamma~^3He}_{y~x}$
    for two-body $^3$He
  photodisintegration $\vec {^3He}(\vec {\gamma},p)d$ at $E_{\gamma}=15$~MeV
(left column) and $E_{\gamma}=30$~MeV (right column).
Lines are the same as in Fig.\ref{fig1}.
\label{fig3}}
\end{figure}

In Figs. \ref{fig1}-\ref{fig3} we show results for two-body photodisintegration
of $^3$He. As shown in the previous section, from the set of 15 possible
polarization observables for $\vec {\gamma} + \vec {^3He} \to p + d$ process
given by Eq. (\ref{eq1.6}) only two analyzing powers: $A^{\gamma}_x$ and
$A^{^3He}_y$,  as well as 5 spin correlation coefficients: $C^{\gamma,{^3}He}_{c,z}$,
 $C^{\gamma,{^3}He}_{c,x}$, $C^{\gamma,{^3}He}_{x,y}$, $C^{\gamma,{^3}He}_{y,z}$, and
$C^{\gamma,{^3}He}_{y,x}$, do not vanish. Two observables, namely the unpolarized
cross section $\frac {d\sigma} {d\Omega}$
 and photon analyzing power $A^{\gamma}_x$ show
  only small sensitivity to the treatment of the two-body contributions
  to the current  operator, and predictions obtained with meson exchanges
  and Siegert theorem are close to each other.  Changing the photon energy from
  $E_{\gamma}=15$~MeV to $30$~MeV diminishes the cross section by a factor of
  $\approx 3$. The  analyzing power $A^{\gamma}_x$ is, especially at the lower energy
  $E_{\gamma}=15$~MeV, large and approaches a value close to 1
   in a wide range of proton angles.
  The $^3$He analyzing power $A^{{^3}He}_y$  reaches at
  $E_{\gamma}=30$~MeV the value of $\approx 0.2$ and reveals at both energies a
  similar and large sensitivity to the current operator.

Among the five nonvanishing spin correlation coefficients (see Figs. \ref{fig2} and
  \ref{fig3}) $C^{\gamma,{^3}He}_{y,z}$ is small and shows only moderate
  sensitivity to the current operator. The spin correlation
  $C^{\gamma,{^3}He}_{c,z}$ (Fig. \ref{fig2}) at $15$~MeV takes large values
  only at very forward and backward angles, being otherwise small.
  With increasing energy its sensitivity to the current grows.
  Spin correlations  $C^{\gamma,{^3}He}_{c,x}$, $C^{\gamma,{^3}He}_{x,y}$, and
$C^{\gamma,{^3}He}_{y,x}$ behave similarly, being  comparable in
  magnitude and showing quite large sensitivity to the underlying current.

\begin{figure}[htbp] 
\includegraphics[scale=0.7]{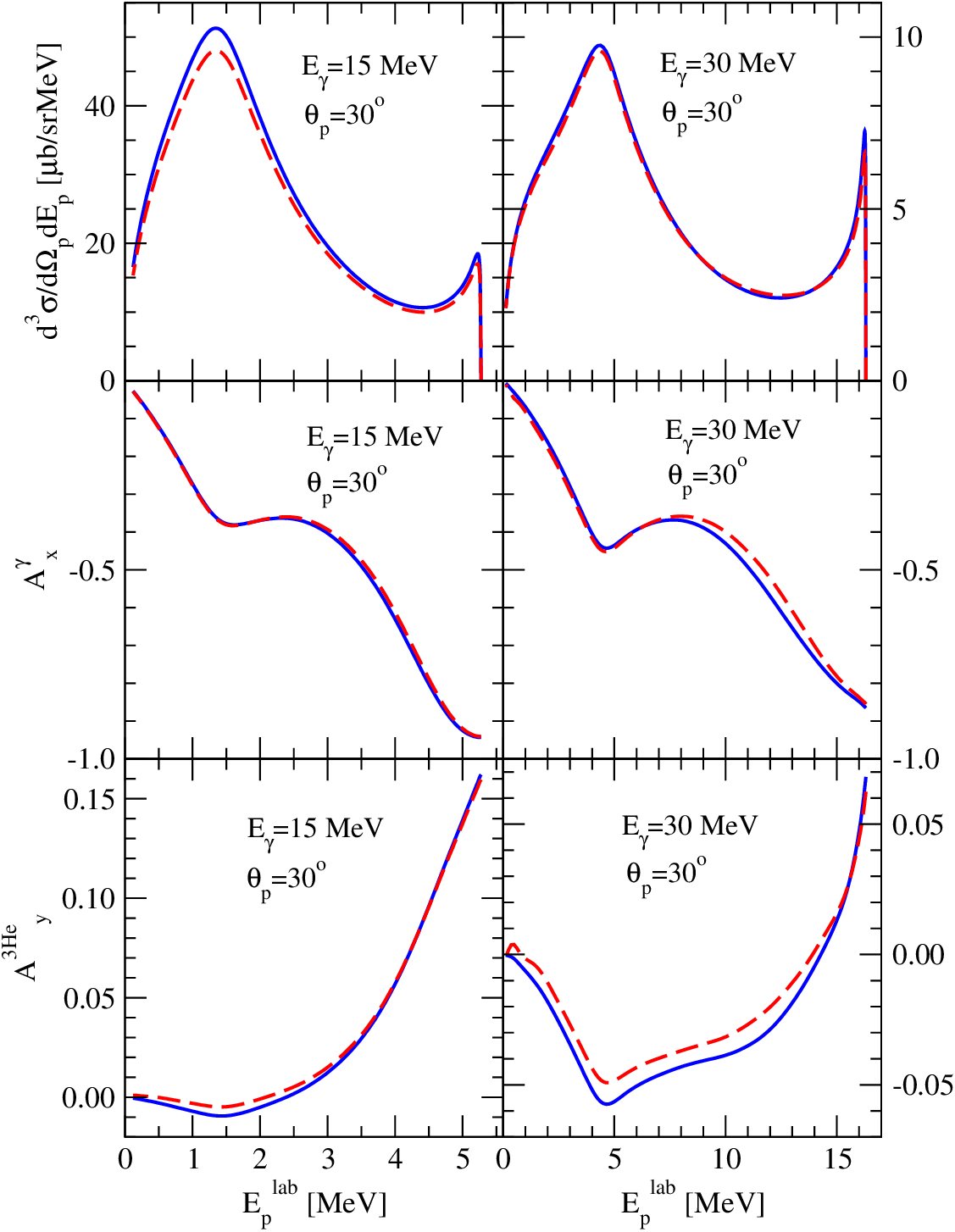}
\caption{(Color online) 
 The unpolarized cross section and the analyzing powers: $A^{\gamma}_x$ and 
  $A^{^3He}_y$, for semi-inclusive  three-body $^3$He
  photodisintegration $\vec {^3He}(\vec {\gamma},p)np$ at $E_{\gamma}=15$~MeV
  (left column) and $E_{\gamma}=30$~MeV (right column) as a function of
  the laboratory energy $E_p^{lab}$ of the outgoing proton detected at lab. angle
  $\theta_p=30^o$. 
  Lines are the same as in Fig.\ref{fig1}.
\label{fig4}}
\end{figure}

\begin{figure}[htbp] 
\includegraphics[scale=0.7]{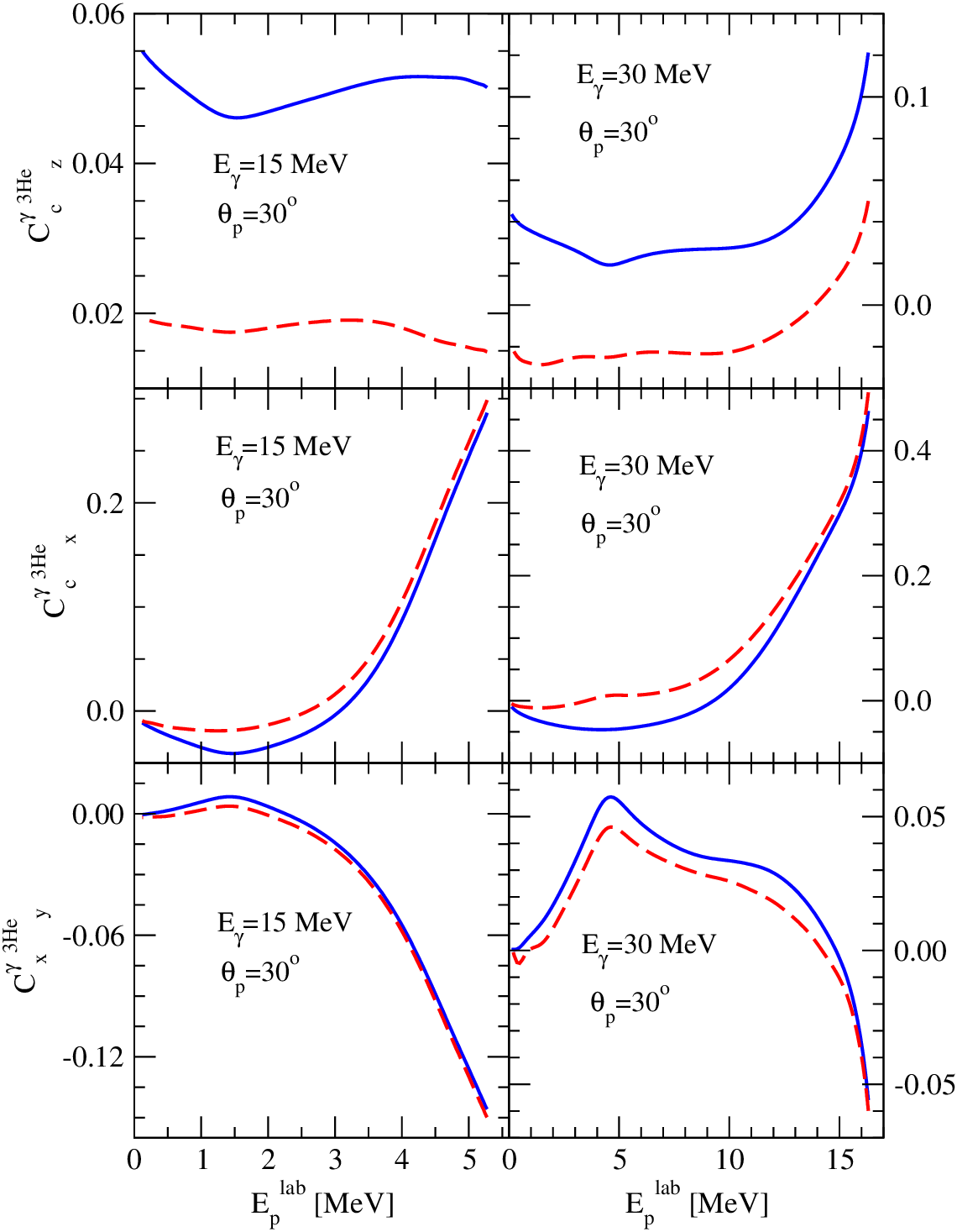}
\caption{(Color online) 
  The same as in Fig.\ref{fig4} but for spin correlation coefficients
   $C^{\gamma-^3He}_{c-z}$, $C^{\gamma-^3He}_{c-x}$, and 
  $C^{\gamma-^3He}_{x-y}$.
  Lines are the same as in Fig.\ref{fig1}.
\label{fig5}}
\end{figure}

\begin{figure}[htbp] 
\includegraphics[scale=0.7]{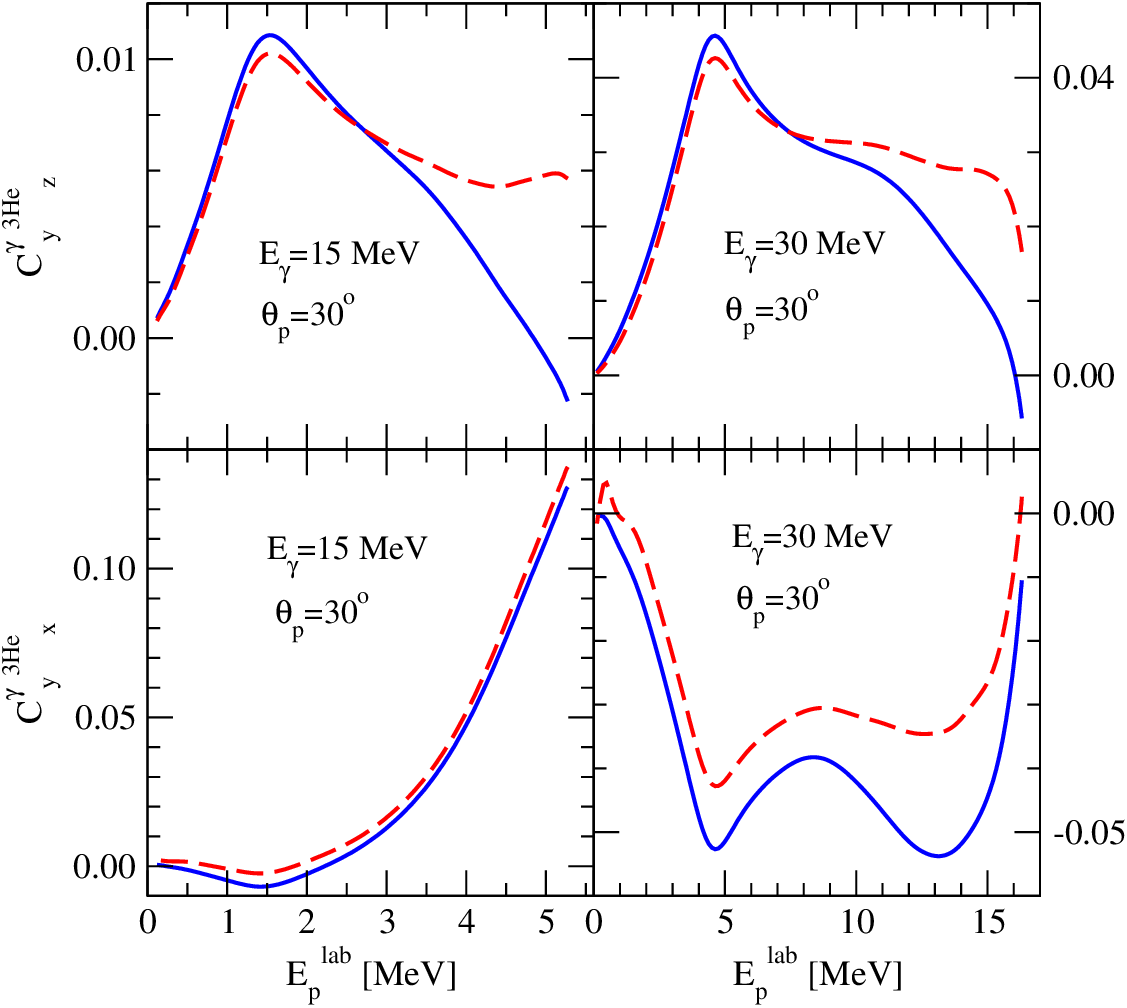}
\caption{(Color online) 
  The same as in Fig.\ref{fig4} but for spin correlation coefficients
   $C^{\gamma~^3He}_{y~z}$ and  $C^{\gamma~^3He}_{y~x}$. 
  Lines are the same as in Fig.\ref{fig1}.
\label{fig6}}
\end{figure}

For semi-inclusive 3-body photodisintegration  $\vec {{^3}He}$($\vec {\gamma}$,p)np
the set of nonvanishing spin observables is the same
 as for the two-body fragmentation. We show them in Figs. \ref{fig4}-\ref{fig6} 
together with the unpolarized cross section as a function of the outgoing
laboratory energy of the detected proton at a laboratory proton 
angle $\theta_p=30^{\circ}$ . The unpolarized cross
section $\frac {d{^3}\sigma}  {d{\Omega}_pdE_p}$ shows a characteristic peak at
the maximum energy of the outgoing proton caused by a final state interaction in
the state $^1S_0$ of the undetected proton-neutron pair. The cross section
drops with increasing photon energy, being reduced by a
 factor of $\approx 4$ when changing the photon
energy from $15$ to $30$~MeV. The unpolarized cross section and the photon
 analyzing power   $A^{\gamma}_x$ shows only small
 sensitivity to the current, contrary to the spin correlation coefficients,
 which exhibit, with exceptions of $C^{\gamma,{^3}He}_{x,y}$ and $C^{\gamma,{^3}He}_{y,x}$
 at E$_\gamma$=15~MeV, quite large differences along the outgoing proton energy
 when  changing the type of two-body contributions to the current operator.
For most of the presented observables that sensitivity seems to be similar 
at both photons energies.

Three-body fragmentation of $^3$He caused by the interaction with incoming
photon offers a unique possibility to reach a specific kinematically
complete geometries of  three outgoing nucleons identical to
those populated in a kinematically complete nucleon-deuteron (Nd) breakup.
Comparison of data in such kinematically complete configurations, reached
either by pure hadronic interactions in Nd breakup or by electromagnetic
and hadronic interactions in three-body fragmentation of $^3$He, would be 
 very  interesting due to unresolved problems found for some kinematically
complete measurements e.g. in the symmetric space-star (SST)
 geometry \cite{din1}.
Data for such geometries gained from complimentary Nd and three-body $^3$He
photodisintegration and their comparison to theory could shed some light
on the existing discrepancies. In Tab.~\ref{tab1} we compare the incoming photon
laboratory energy $E_{\gamma}$ of the $^3$He photodisintegration with the
laboratory kinetic energy $E^{lab}$ of the incoming nucleon in the corresponding
proton-induced deuteron breakup reaction, both processes
 having the same 3N system
 center-of-mass kinetic energy $E^{c.m.}$. 

The kinematically complete geometries of three outgoing nucleons can
be defined for three-body photodisintegration  of $^3$He
 analogously to their Nd breakup
counterparts. For example for the final state configuration (FSI) 
the  condition is the equality of  momenta of two interacting
nucleons in the final scattering state. For the SST three outgoing
nucleons must 
move in their c.m. system  with equal momentum magnitudes in a
 plane perpendicular
to the incoming photon momentum and
the relative angle between momenta is 120$^{\circ}$ like in the ``Mercedes star'' logo.

\begin{figure}[htbp] 
\includegraphics[scale=0.7]{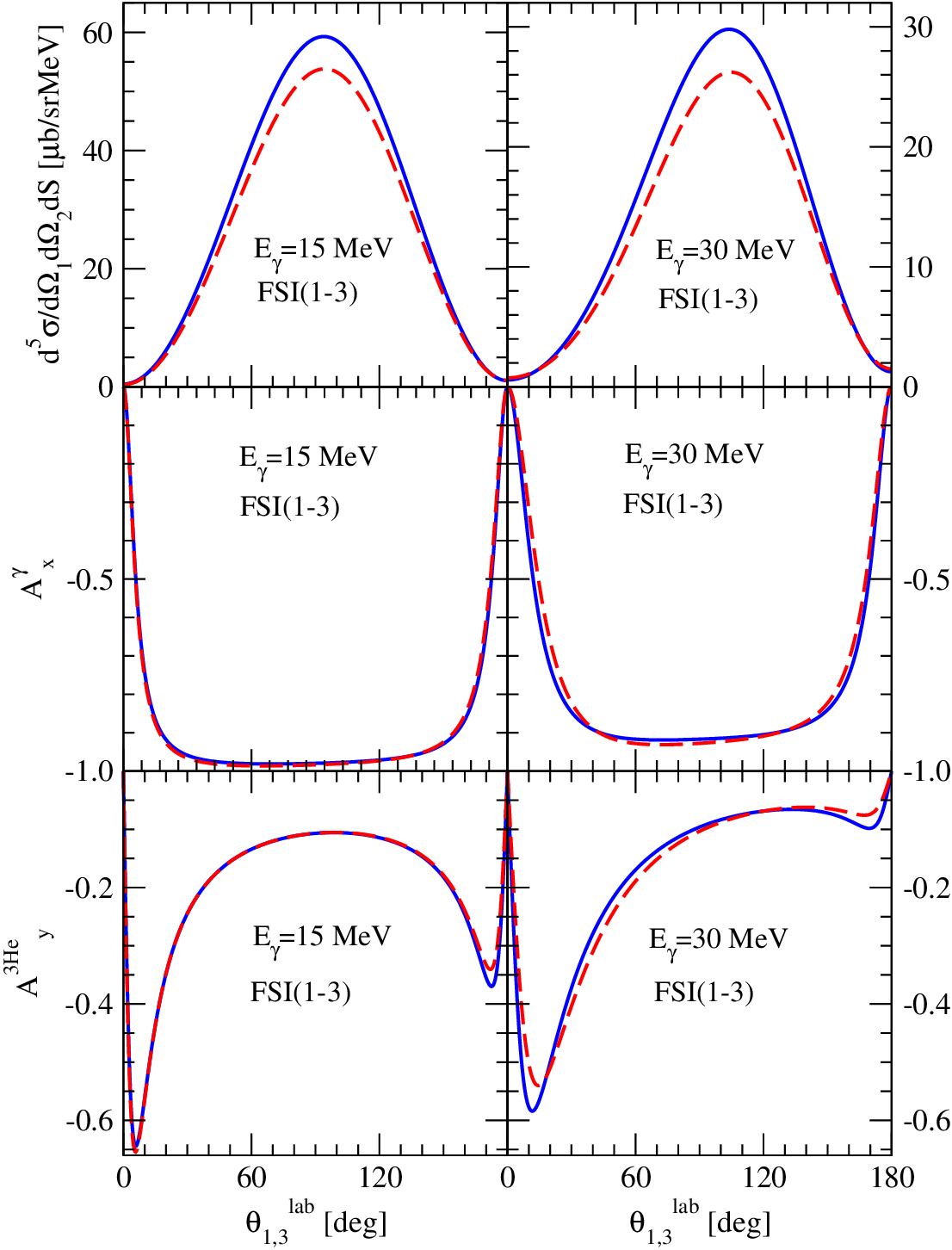}
\caption{(Color online) 
  The unpolarized cross section  $\frac {d^5\sigma} {d\Omega_1d\Omega_2dS}$
  and the analyzing powers: $A^{\gamma}_x$ and 
  $A^{^3He}_y$, for exclusive  three-body $^3$He
  photodisintegration $\vec {^3He}(\vec {\gamma},pp)n$ at $E_{\gamma}=15$~MeV
  (left column) and $E_{\gamma}=30$~MeV (right column), for kinematically
  complete final state interaction configuration, where nucleons 1 and 3
  (proton and neutron, respectively) have the same momenta
  $\vec p_1=\vec p_3$. The observables are shown as a function of the laboratory
  angle $\theta_1^{lab}=\theta_3^{lab}$ at which that configuration is produced.  
Lines are the same as in Fig.\ref{fig1}.
\label{fig7}}
\end{figure}

\begin{figure}[htbp] 
\includegraphics[scale=0.7]{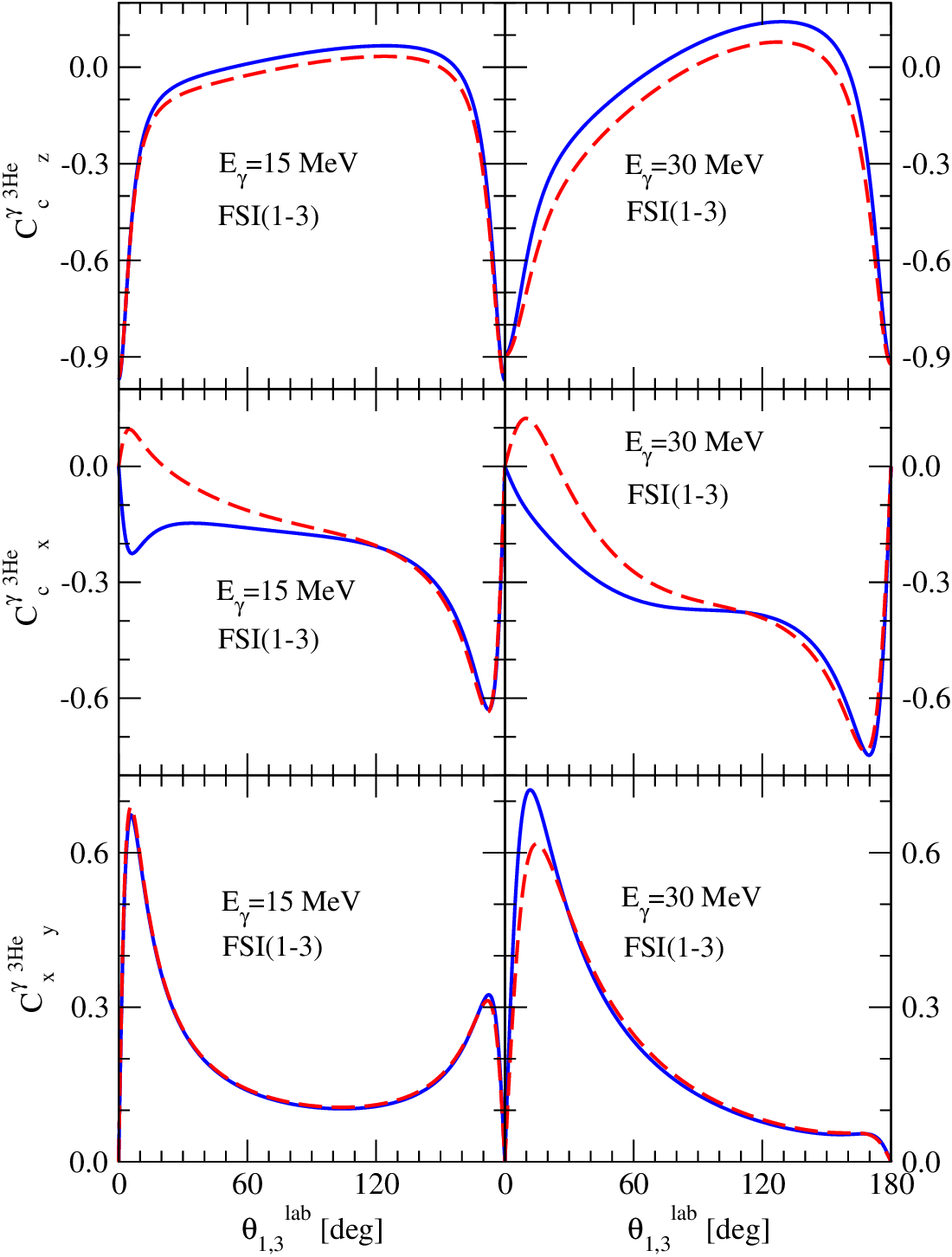}
\caption{(Color online) 
  The same as in Fig.\ref{fig7} but for spin correlation coefficients
   $C^{\gamma-^3He}_{c-z}$, $C^{\gamma-^3He}_{c-x}$, and 
  $C^{\gamma-^3He}_{x-y}$.
  Lines are the same as in Fig.\ref{fig1}.
\label{fig8}}
\end{figure}

\begin{figure}[htbp] 
\includegraphics[scale=0.7]{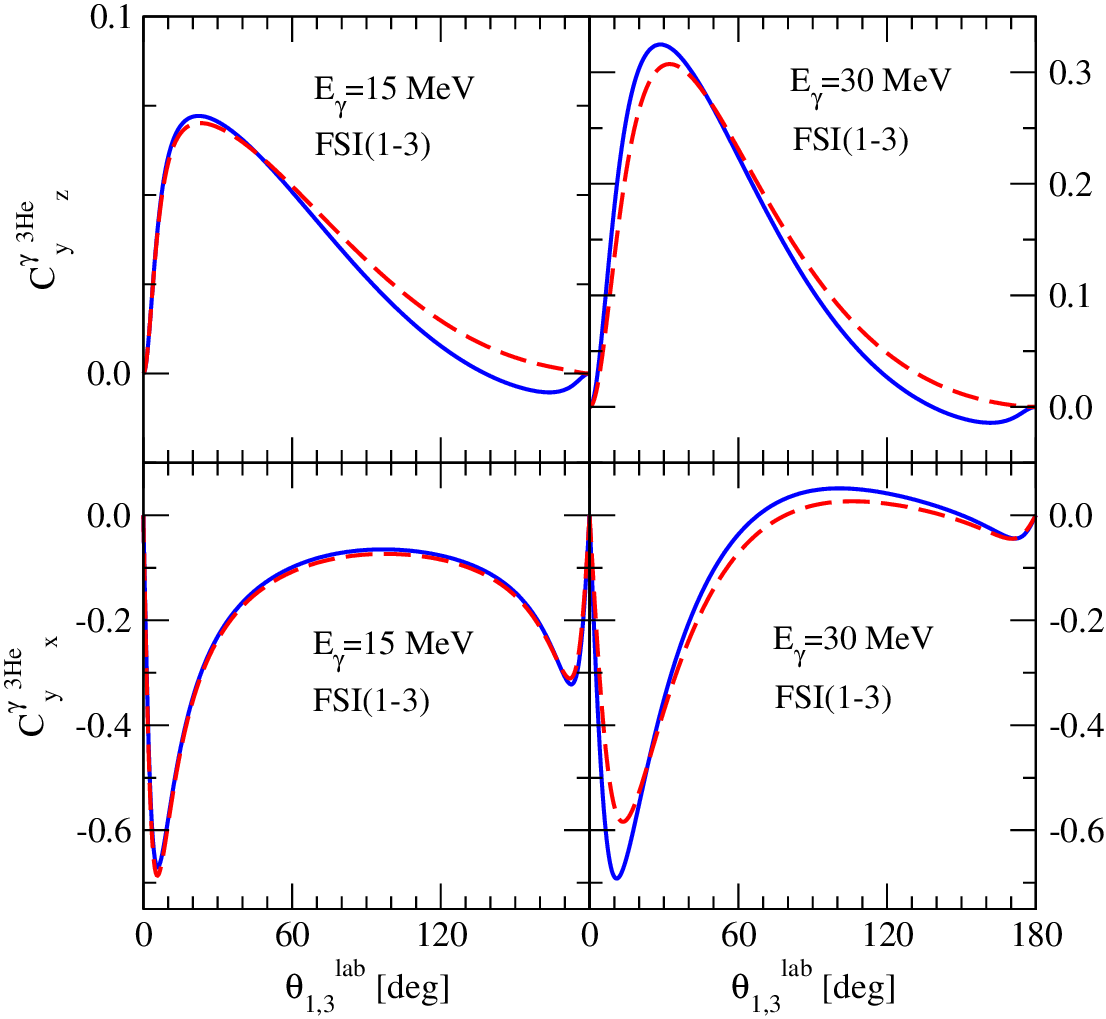}
\caption{(Color online) 
  The same as in Fig.\ref{fig7} but for spin correlation coefficients
   $C^{\gamma~^3He}_{y~z}$ and  $C^{\gamma~^3He}_{y~x}$. 
  Lines are the same as in Fig.\ref{fig1}.
\label{fig9}}
\end{figure}

In the case of the final state interaction the condition of momentum equality of two
nucleons permits finding, at each laboratory direction of outgoing nucleon, 
such a FSI geometry in which one from remaining two
 nucleons have the same momentum.  
In Figs. \ref{fig7}-\ref{fig9} we show nonvanishing observables for
exclusive three-body photodisintegration of $^3$He:
$\vec {{^3}He}(\vec {\gamma},pp)n$,  leading to a kinematically
  complete final state interaction configuration in which nucleons 1 and 3
  (in this case proton and neutron, respectively) have the same momenta
  $\vec p_1=\vec p_3$. The observables are shown as a function of a production
  angle  $\theta_1^{lab}=\theta_3^{lab}$ of such a FSI(1-3) configuration.
   The unpolarized
  cross section $\frac {d{^5}\sigma}  {d{\Omega}_1d{\Omega}_2dS}$ shown
  in Fig. \ref{fig7} reaches largest values for FSI(1-3) configurations
  produced  at angles around $\theta_1^{lab} \approx 100^o$. Similarly to
  inclusive breakup, increasing the photon energy from $15$ to $30$~MeV diminishes
  the cross section by a factor of $ \approx 2$. The FSI cross section at
  both energies is quite sensitive to the underlying current.

The FSI analyzing powers  $A^{\gamma}_x$ and  $A^{{^3}He}_y$ show
some slight sensitivity to the current operator
 only
  at the higher energy $E_{\gamma}=30$~MeV. 
Interestingly enough, they are quite
  large in a wide range of FSI production angles.

FSI spin correlation coefficients offer more sensitivity to the current, 
  particularly $C^{\gamma,{^3}He}_{c,z}$ at $E_{\gamma}=30$~MeV
  and $C^{\gamma,{^3}He}_{c,x}$ at both energies. Also FSI spin correlation
  coefficients take large values in quite large regions of
   the FSI production angles.

Among numerous kinematically complete configurations of the Nd breakup 
reaction the SST configuration has attracted a special  attention.
 The  cross  section  for
that geometry is very stable with respect to the underlying dynamics and 
dominated by the S-waves \cite{din1}. At low energies theoretical predictions
deviate significantly from the available SST
 data \cite{koln13,sst_tunl,sagarasst}. 
The possibility to reach that geometry through $^3$He three-body
  photodisintegration
 $^3$He($\gamma$,pp)n would help to shed some light on this problem.

\begin{table}
\begin{tabular}{|c|c|c|c|}
\hline
$E_{\gamma}$~[MeV] & $E^{c.m.}$~[MeV] & $E^{lab}$~[MeV] & $p^{lab}$~[MeV/c] \\
\hline
15.0  &  7.2   &  14.2    &  164.2   \\
20.0  &  12.2  &  21.7    &  203.2   \\
30.0  &  22.1  &  36.7    &  265.2   \\
\hline
\end{tabular}
\caption{The initial photon laboratory energy $E_{\gamma}$, the consequent
  center-of-mass kinetic energy of the $ppn$ system $E^{c.m.}$
  and the incident proton laboratory kinetic energy $E^{lab}$ together
  with its momentum $p^{lab}$ for the corresponding proton-induced
   deuteron breakup reaction with the same $E^{c.m.}$.}
\label{tab1}
\end{table}

\begin{figure}[htbp] 
\includegraphics[scale=0.7]{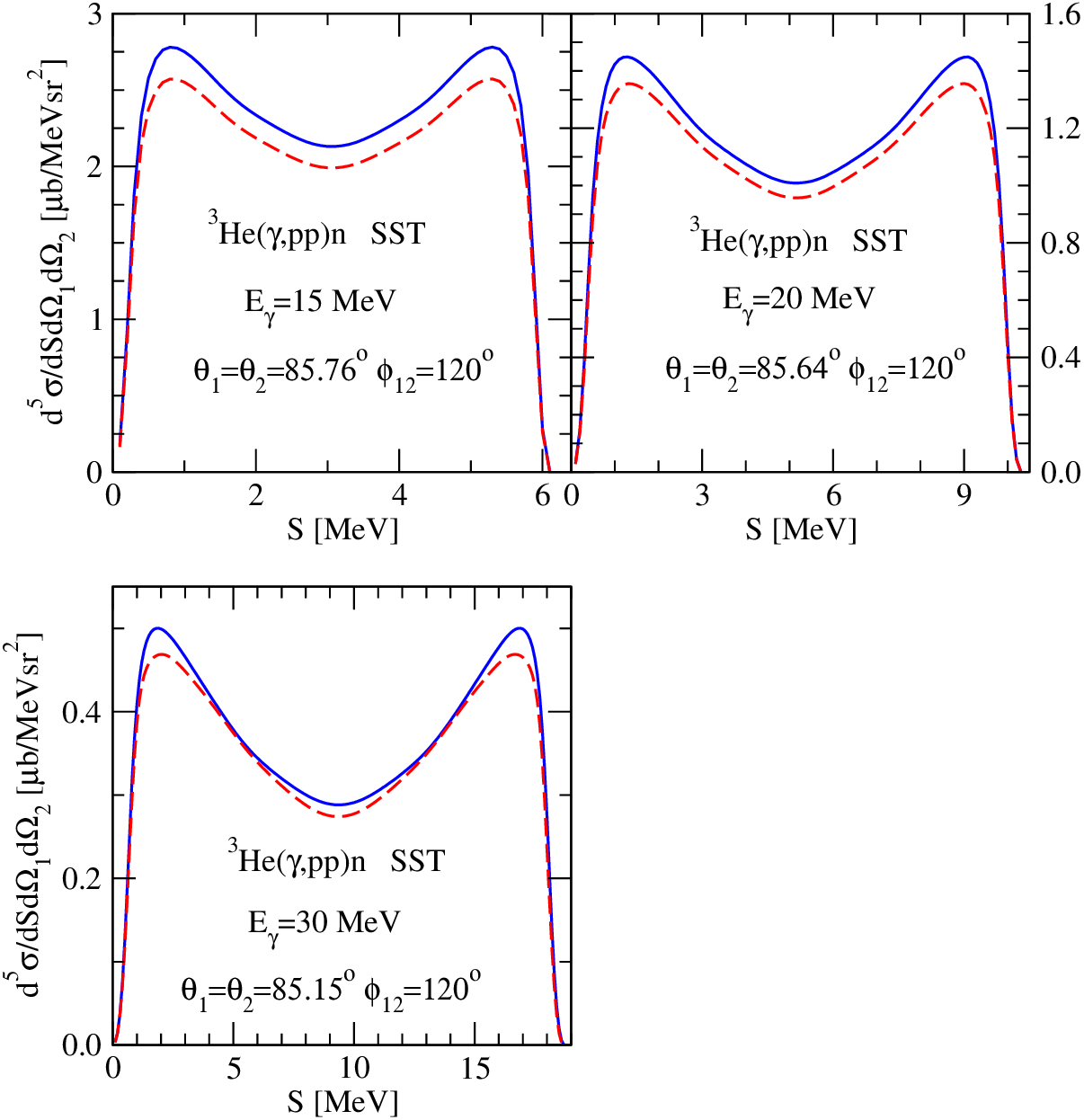}
\caption{(Color online) 
 Unpolarized cross section for exclusive  three-body $^3$He
 photodisintegration $\vec {^3He}(\vec {\gamma},pp)n$ at $E_{\gamma}=15$,
 $20$, and $30$~MeV, in a kinematically
 complete space star geometry, shown as a function of arc-length of the S-curve 
 in the plane of  laboratory energies of nucleons 1 and 2. 
 Lines are the same as in Fig.\ref{fig1}.
\label{fig10}}
\end{figure}

In Fig. \ref{fig10} we show the SST cross section
$\frac {d{^5}\sigma}  {d{\Omega}_1d{\Omega}_2dS}$  at
three incoming photon energies
 $E_{\gamma}=15$, $20$, and $30$~MeV as a function of
an arc-length of the S-curve. This curve,
in the plane determined by the laboratory energies
of two detected
 protons $E_1 - E_2$, contains all the kinematically allowed events.
The SST
condition is exactly fulfilled at a central part of the S-curve at each
photon energy. The SST cross section observed
 along the S-curve reveals only a weak
sensitivity to the
underlying current  and the cross section drops quite rapidly
with increasing photon energy, changing from
$\frac {d^5\sigma} {d{\Omega}_1d{\Omega}_2dS} \approx 2$~$\frac {mb} {sr^2MeV}$
at $E_{\gamma}=15$~MeV to
$\frac {d^5\sigma} {d{\Omega}_1d{\Omega}_2dS} \approx 0.3$~
 $\frac {mb} {sr^2MeV}$ at $E_{\gamma}=30$~MeV.

\section{Summary and Conclusions}
\label{summary}

Recent advances in high intensity polarized photon beams and polarized
 $^3$He targets increased  the number of observables 
to be measured in two- and three-body
 photodisintegration  of $^3$He. In addition
 to the unpolarized cross section also measurements of photon and $^3$He
 analyzing powers as well as
 spin correlation coefficients are feasible. That possibility together
 with promising results 
 achieved in derivation, in the framework of chiral perturbation theory, of
 consistent
 two- and three-nucleon forces as well as electro-weak currents, allows
for comprehensive testing the
 chiral dynamics not only in pure hadronic systems but also 
in processes induced by the interaction of external electro-weak probes with
 nuclear systems.

Photodisintegration of polarized $^3$He by polarized photon
 not only provides a rich choice
 of observables to be measured but, due to the availability of rigorous Faddeev
 techniques for solving the
 corresponding equations, allows one to compare such data with
 exact theoretical predictions for
spin-dependent observables, thus extending the testing ground  for nuclear dynamics.

We investigated the observables in two- and three-body fragmentation
of $^3$He with respect to their
sensitivity to the underlying current by comparing results with two
different treatments of two-body
contributions, namely by taking them as direct meson exchanges or
treating them by using Siegert theorem.

For two-body fragmentation we found that the unpolarized cross section
and the photon analyzing
power $A^{\gamma}_x$ are practically insensitive to the underlying current.
Large
values of $A^{\gamma}_x$ point to the feasibility of its measurement. Both
these observables would be valuable for future testing of chiral dynamics. 
The sensitivity to the current operator of the $^3$He analyzing power $A^{{^3}He}_y$ and
the spin correlation coefficients predisposes them to test the current operator.

For semi-inclusive three-body fragmentation we found a similar behavior.
 While the unpolarized cross section
 and photon analyzing power $A^{\gamma}_x$ show only a slight sensitivity to
 the current, the 
 $^3$He analyzing power and spin correlation coefficients reveal sensitivity
 to the current
 practically at all the energies of the detected nucleon.

 From the rich phase-space of the exclusive three-body $^3$He fragmentation
  we investigated
 only the geometry of the final-state interaction
 and the symmetric space-star configuration. For FSI we found a sensitivity
 of the cross section to the
 underlying current while the analyzing powers show only very slight sensitivity.
 Among spin correlation
 coefficients the largest sensitivity is visible in $C^{\gamma,{^3}He}_{c,x}$.
 The sensitivity of the unpolarized SST cross section to the underlying
 current gets reduced with the increasing photon energy. 

 Summarizing, measurements of $\vec {{^3}He} + \vec {\gamma}$
 observables for both two- and
 three-body $^3$He fragmentation seem feasible and such data would
  provide a valuable
test for our understanding of electromagnetic processes, especially 
in the context of expected results based on chiral dynamics. We hope this work will 
guide preparations of new generation precise experiments on $^3$He photodisintegration.

\begin{acknowledgments}
This study has been performed within Low Energy Nuclear Physics
International Collaboration (LENPIC) project and 
was  supported by the Polish National Science Center 
 under Grants No. 2016/22/M/ST2/00173 and 2016/21/D/ST2/01120. 
 The numerical calculations were performed on the 
 supercomputer cluster of the JSC, J\"ulich, Germany.
\end{acknowledgments}



%
%
%
%
\newpage
\newpage
\newpage
\newpage
\newpage
\newpage
\newpage
\newpage
\newpage
\newpage

\end{document}